\documentstyle[12pt]{article} \hoffset -0.5in \textwidth 6.5in \textheight
9.00in  \parskip 7pt \openup2.5\jot \parindent=0.5in
\newskip\humongous \humongous=0pt plus 1000pt minus 1000pt
\def\caja{\mathsurround=0pt}
\def\eqalign#1{\,\vcenter{\openup1\jot \caja
        \ialign{\strut \hfil$\displaystyle{##}$&$
        \displaystyle{{}##}$\hfil\crcr#1\crcr}}\,}
\newif\ifdtup

\def\eqright #1\cr{\noalign{\hfill$\displaystyle{{}#1}$}}
\def\eqleft #1\cr{\noalign{\noindent$\displaystyle{{}#1}$\hfill}}
\def\oldreffmt#1{\rlap{[#1]} \hbox to 2\parindent{}}

\def\figfmt#1{\rlap{Figure {#1}} \hbox to 1in{}}

\def\auto{\eqno(\refstepcounter{equation}\theequation)}
\def\begineq #1\endeq{$$ \refstepcounter{equation}\eqalign{#1}\eqno
        (\theequation) $$}
\def\contlimit{\,{\hbox{$\longrightarrow$}\kern-1.8em\lower1ex
\hbox{${\scriptstyle (a\rightarrow0)}$}}\,}
\def\centeron#1#2{{\setbox0=\hbox{#1}\setbox1=\hbox{#2}\ifdim
\wd1>\wd0\kern.5\wd1\kern-.5\wd0\fi
\copy0\kern-.5\wd0\kern-.5\wd1\copy1\ifdim\wd0>\wd1
\kern.5\wd0\kern-.5\wd1\fi}}
\def\centerover#1#2{\centeron{#1}{\setbox0=\hbox{#1}\setbox
1=\hbox{#2}\raise\ht0\hbox{\raise\dp1\hbox{\copy1}}}}
\def\centerunder#1#2{\centeron{#1}{\setbox0=\hbox{#1}\setbox
1=\hbox{#2}\lower\dp0\hbox{\lower\ht1\hbox{\copy1}}}}
\def\lsim{\;\centeron{\raise.35ex\hbox{$<$}}{\lower.65ex\hbox
{$\sim$}}\;}
\def\gsim{\;\centeron{\raise.35ex\hbox{$>$}}{\lower.65ex\hbox
{$\sim$}}\;}
\def\super#1{\ifmmode \hbox{\textsuper{#1}}\else\textsuper{#1}\fi}
\def\textsuper#1{\newcount\holdspacefactor\holdspacefactor=\spacefactor
$^{#1}$\spacefactor=\holdspacefactor}
\def\getcite#1,{\advance\citenumber by1
\ifnum\citenumber=1
\ref{#1}\let\next=\getcite\else\ifx#1@\let\next=\relax
\else ,\ref{#1}\let\next=\getcite\fi\fi\next}
\def\upon #1/#2 {{\textstyle{#1\over #2}}}
\relax
\renewcommand{\thefootnote}{\fnsymbol{footnote}}

\def\til#1{\centeron{\hbox{$#1$}}{\lower 2ex\hbox{$\char'176$}}}
\def\tild#1{\centeron{\hbox{$\,#1$}}{\lower 2.5ex\hbox{$\char'176$}}}
\def\sumtil{\centeron{\hbox{$\displaystyle\sum$}}{\lower
-1.5ex\hbox{$\widetilde{\phantom{xx}}$}}}

\def\pom{{\rm P\kern -0.53em\llap I\,}}
\def\spom{{\rm P\kern -0.36em\llap \small I\,}}
\def\sspom{{\rm P\kern -0.33em\llap \footnotesize I\,}}

\begin{document} \begin{titlepage}
\rightline{\vbox{\halign{&#\hfil\cr
&ANL-HEP-PR-94-23\cr
&\today\cr}}}
\vspace{0.25in}
\begin{center}

{\Large\bf
THE $O(g^4)$ LIPATOV KERNELS}
\medskip

Alan R. White\footnote{Work
supported by the U.S. Department of
Energy, Division of High Energy Physics, Contract\newline W-31-109-ENG-38}
\\ \smallskip
High Energy Physics Division\\Argonne National
Laboratory\\Argonne, IL 60439\\ \end{center}

\begin{abstract}
Leading plus next-to leading log results for the Regge limit of massless
Yang-Mills theories are reproduced by reggeon diagrams in which the Regge
slope $\alpha' \to 0$ and reggeon amplitudes satisfy Ward identity
constraints at zero transverse momentum. Using reggeon unitarity together
with multiple discontinuity theory a complete set of such diagrams can be
constructed. The resulting two-two, one-three and two-four kernels which
generalise the Lipatov equation at $O(g^4)$ are determined uniquely.

\end{abstract}

\renewcommand{\thefootnote}{\arabic{footnote}} \end{titlepage}

The Regge limit of $QCD$ has been the focus of much theoretical interest
recently. In one development it has been shown that the leading-log
perturbative calculations\cite{lip,bfkl,cl} can be reproduced by a
two-dimensional effective theory\cite{kls,vv} and it is hoped that this can
be used in an $s$-channel unitarisation procedure. Experimental results from
HERA, CERN, and the Tevatron Collider, on the small-x behavior of structure
functions and other ``hard diffractive'' phenomena have also provided a
stimulus for theoretical developments. At the center of much of the
discussion is the Lipatov equation\cite{lip} which is applied, in
particular, to determine the leading log small-x evolution of parton
distributions. Also, the Pomeron which emerges from the Lipatov equation is
the general focus of much theoretical activity and of comparisons with
experiment. Consequently, there has been considerable emphasis on the need to
derive non-leading log corrections to the equation.

In this letter we will report results which demonstrate the power of
$t$-channel, rather than $s$-channel, unitarity in determining corrections to
the leading-log results. We give explicitly what we argue to be the full
$O(g^4)$ modification of the Lipatov two-two gluon kernel - which is
$O(g^2)$ in leading-order - together with corresponding one-three and
two-four kernels which appear for the first time at $O(g^4)$. We believe
that, with sufficient effort, our results can be straightforwardly extended
to even higher orders. However, we should emphasize that our formalism
accounts only for the contributions of ``nonsense'' exchanged gluon states
that reduce to transverse momentum integrals. Such states give a complete,
self-consistent, description of the infra-red transverse momentum dependence
of the theory in the Regge limit. The ``running'' of the gauge coupling $g$,
due to short-distance renormalization, will not appear explicitly nor will
the dependence of the theory on any other states.

We recall that it is $t$-channel unitarity, expressed via {\it reggeon
unitarity}\cite{gpt,arw1}, which requires that the Regge limit be described
by an effective two-dimensional theory in the form of {\it reggeon
diagrams}. The reggeon diagrams describing the leading plus next-to-leading
log results have been known for a long time\cite{reg}, together with the
interpretation of the Lipatov equation as a Bethe-Salpeter equation for
reggeon amplitudes. In a companion paper\cite{arw3} we show how a
combination of reggeon unitarity and multiple discontinuity theory can be
used to build the leading plus next-to-leading diagrams into a complete set.
In this paper we will only briefly outline the construction procedure. We
will concentrate on those reggeon diagrams which give the new $O(g^4)$
kernels. We will show that these kernels are uniquely determined by the
combination of t-channel unitarity, implemented via reggeon diagrams,
with our implementation of gauge invariance. We impose gauge invariance by
two conditions. First, that reggeon amplitudes vanish when the transverse
momentum of any reggeon goes to zero, and secondly that ($t$-channel) color
zero amplitudes are infra-red finite - as integral kernels.

Imposing the vanishing of reggeon amplitudes at zero tranverse momentum is
directly equivalent to imposing the defining Ward identities of the
theory\cite{gth}. The reggeon amplitudes we construct are necessarily
gauge-invariant since they can each be used to describe Regge limits of
multiparticle S-Matrix elements. However, as the transverse momentum of a
reggeon vanishes it can be identified with an elementary gluon carrying {\it
zero four-momentum}. If the remainder of the reggeon amplitude under
discussion is embedded in an on-shell S-Matrix amplitude as illustrated in
Fig.~1(a), we obtain the zero momentum limit of the amplitude for an
off-shell gluon to couple in to the S-Matrix element. This amplitude
satisfies a Ward identity\cite{gth} as shown in Fig.~1(b). If there are no
internal infra-red divergences occurring explicitly at zero transverse
momentum (as will be the case in the absence of massless
fermions\cite{arw2}), then, as illustrated, this identity requires the
amplitude to vanish.

There is no scale in our (perturbative) formulae, just a bare,
dimensionless, gauge coupling. As a result individual reggeon diagrams do
contain infra-red divergences. We have discussed at length in \cite{arw2}, the
dynamical implications of these divergences for the ``non-perturbative''
solution of the theory. In this paper we will not discuss this point and
instead give only the infra-red finite combinations of diagrams that provide
generalised Lipatov kernels.

The existing results which go beyond the $O(g^2)$ approximation for the
Lipatov kernel are essentially all due to Bartels\cite{reg,bar}. By exploiting
$s$-channel unitarity, together with the analytic structure of multi-Regge
production amplitudes, Bartels has formulated a general construction
procedure for reggeon diagrams. As a result, a large class of reggeon diagrams
has been constructed and many elements of the reggeon kernels we give obtained.
We will briefly discuss at the end of the paper how we believe our full
kernels emerge in this construction. Our method is related to that of
Bartels in that we exploit heavily basic Regge properties of analyticity and
unitarity\cite{arw1}. However, we do not calculate production amplitudes at
all, but rather input all the constraints of multiparticle discontinuity
theory directly at the reggeon diagram level. We believe this enables us to
go systematically further in the construction of diagrams.

Our method has some parallel with the well-known Bern-Kosower\cite{bk}
technique for calculating multi-gluon amplitudes from string amplitudes. That
is we introduce a Regge slope $\alpha'$ for gluons, but instead of beginning
with string amplitudes, we begin with the general class of reggeon diagrams
for interacting odd-signature reggeons. We build in the gauge group via the
reggeon interactions. (In this paper, however, we ignore group factors
almost entirely since they essentially disappear from the color zero
amplitudes that are the focus of our discussion.) A direct manifestation
of our imposition of gauge invariance is that all elementary reggeon
couplings, which in principle could be independent parameters, are
determined in terms of one parameter - the triple reggeon coupling.
Perturbation theory is recovered if the triple reggeon coupling is
identified as the gauge coupling $g$ scaled by $\sqrt{\alpha'}$ and the
limit $\alpha' \to 0$ is taken with $g$ kept fixed.

We can illustrate the essence of the method developed in \cite{arw3} by
outlining the derivation of the Lipatov equation from reggeon diagrams. A
general reggeon diagram (with one or more cuts through it) contains a
reggeon propagator $$
\eqalign { \Gamma_n = {1
\over \left(E-\sum^n_{i=1} \alpha'k^2_i\right)}}
$$
for each $n$-reggeon intermediate state, a particle pole or ``signature
factor''
$$
\eqalign{\sigma(k^2)~=~{1 \over \alpha'k^2}}\auto
$$
for each uncut reggeon line, together with elementary (cut) reggeon vertices
- which are simple functions of transverse momenta. The transverse momentum
loop integrals are, of course, two-dimensional with momentum conservation
imposed. The triple reggeon vertex contains a ``nonsense'' zero which
manifests itself in various forms i.e.
$$
\eqalign{ \Gamma_{1,2}~~&\sim ~~g~\sqrt{\alpha'}~[E - \alpha'k_1^2 -
\alpha'k_2^2]\cr
&\sim ~~~[k^2 - k_1^2 -
k_2^2] ~~~~~~~~~~when~~~E~=~\alpha'k^2\cr
&\sim~~~ k^2 ~~~~~~~~~~~~~~~~~~~~~~when~~~k_1^2=k_2^2=0 }\auto
$$

The existing perturbative results are given by the simplest reggeon diagrams.
Gluon reggeization comes from the set of ``self-energy'' diagrams and the
importance of constructing discontinuities is actually apparent at
this first stage. In the cut diagrams shown in Fig.~2 it is clear that all
two reggeon propagators are cancelled by triple reggeon nonsense zeroes - in
the first form given above. One zero per loop remains to provide, in the
third form given, the reggeisation. The real part is obtained by adding a
signature factor to the imaginary part obtained from the sum of the cut
diagrams. The same result is {\it not obtained} if we try to construct the
sum of the uncut diagrams directly\cite{arw3}. Defining
$$
\eqalign{ J_1(k^2)~~=~~{1 \over (2\pi)^3}\int d^2q {1 \over
q^2(k-q)^2}}\auto
$$
the sum in Fig.~1 gives (with $g$ appropriately identified) a reggeon
propagator
$$
\eqalign{[E - \alpha'k^2 - g^2~k^2J_1(k^2)]^{-1}
\centerunder{$\longrightarrow$} {\raisebox{-5mm} {$\alpha' \to 0$}}
[E - g^2~k^2J_1(k^2)]^{-1}}\auto
$$
which is the perturbative reggeization result.

To construct multi-loop diagrams we have to proceed loop-by-loop, utilising
multiple discontinuities\cite{arw3}. For the two-loop diagrams shown in
Fig.~3, we first construct triple discontinuities of the triple reggeon
diagrams shown in Fig.~4. From the triple discontinuities, full one-loop
amplitudes can be constructed by adding signature factors. The one-loop
triple reggeon amplitudes can then be integrated with an elementary triple
reggeon vertex to give the (discontinuities of the) diagrams of Fig.~3. In
the triple discontinuities, nonsense zeroes cancel the internal reggeon
propagators and the cuts remove all but two of the internal signature
factors. Hence the diagrams reduce to transverse momentum integrals as
illustrated. The phases of the cut vertices actually determine that the two
triple discontinuities shown are equal but opposite in sign and as a result
the diagrams of Fig.~3 cancel. (If this cancellation did not take place, the
reggeization would be destroyed).

In the {\it color zero}, even-signature, channel the two-loop diagrams are
of the form shown in Fig.~5, where we have introduced a four reggeon
vertex, the determination of which, we will discuss further shortly. The
diagrams are constructed similarly to those of Fig.~3. In all diagrams the
two reggeon propagators remain while the three reggeon propagators are
cancelled by nonsense zeroes. Taking the external couplings to be
$\alpha'g^2$, the sum total of two-loop diagrams can be written in the form
$$
\eqalign{ {g^4 \over (2\pi)^6} &\int {d^2k_1 \over k_1^2} {d^2k_2 \over
k_2^2} \delta^2(k-k_1-k_2)
{1 \over (E-\alpha'k_1^2 - \alpha'k_2^2)}\cr
&\int {d^2k_3 \over k_3^2} {d^2k_4 \over k_4^2} \delta^2(k-k_3-k_4)
{1 \over (E-\alpha'k_3^2 - \alpha'k_4^2)} ~K^{(2)}_{2,2}(k_1,k_2,k_3,k_4)}\auto
$$
where, if we identify the triple reggeon vertex as we have described and
choose the four-reggeon vertex appropriately, $K^{(2)}_{2,2}(k_1,k_2,k_3,k_4)$
is the $O(g^2)$ Lipatov kernel. In particular, if the gauge group is $SU(2)$,
we have
$$
\eqalign{  g^{-2}K^{(2)}_{2,2}(k_1,k_2,k_3,k_4)~=
{}~\sum_{\scriptscriptstyle 1<->2}~
\Biggl(&~(2\pi)^3~k_1^2J_1(k_1^2)k_2^2~\Bigl(k_3^2\delta^2(k_2-k_4)~+~
k_4^2\delta^2(k_2-k_3)\Bigr)~\cr
&-~{k_1^2k_4^2~+~k_2^2k_3^2 \over (k_1-k_3)^2}~
-~(k_1+k_2)^2\Biggr)}\auto
$$
(for $SU(N)$ we simply write $~g^2 \to 2N^{-1}g^2~$). This expression can be
represented by transverse momentum diagrams as in Fig.~6. The notation then
makes transparent the origin of each term in a corresponding reggeon
diagram of the form shown in Fig.~5. The limit $\alpha' \to 0$ of (5) gives
directly the sixth-order perturbative result. Iteration of the construction
procedure to obtain the series of diagrams shown in Fig~7, together with the
resulting Lipatov equation, is straightforward.

Suppose, for the moment, that we know only that the kernel originates from
reggeon diagrams of the three forms shown in Fig.5 and that these give
the three classes of transverse momentum diagrams appearing in
Fig.6. The three classes of diagrams give directly the three terms with
distinct transverse momentum dependence in $K^{(2)}_{2,2}$. We can determine
their relative magnitude as follows.

We first impose the Ward identity constraint
$$
\eqalign { K^{(2)}_{2,2}(k_1,k_2,k_3,k_4)
\centerunder{$\longrightarrow$} {\raisebox{-5mm} {$k_i \to 0$}}
{}~~0~~~i=1,.,4} \auto\label{ginv}
$$
(with $k_1+k_2~=~k_3+k_4~$ imposed, of course). If we consider the limit $k_1
\to 0$ of $K^{(2)}_{2,2}$, with $k_2,k_3,k_4 \neq 0$, the first term in (6)
is eliminated because of the $\delta$-function constraint
$k_2=k_4~(~=>~k_1=k_3)$. The second and third terms give
$$
\eqalign{ ~{k_2^2k_3^2~ \over k_3^2}~+~{k_2^2k_4^2 \over (k_2-k_3)^2}~-~2k_2^2
{}~=~0}\auto
$$
and so this vanishing fixes the relative magnitude of the two terms. Since
the third term is directly given by the four-reggeon vertex it is also clear
that the gauge invariance constraint (\ref{ginv}) determines the magnitude
of the four-reggeon vertex in terms of the three reggeon vertex.

Because it carries ($t$-channel) color-zero, all internal infra-red
divergences of $K^{(2)}_{2,2}$, regarded as an integral kernel, must cancel.
In particular, the divergence as $(k_1-k_3) \to 0$ in the second term must
cancel (after integration) with the divergence of $J(k_1^2)$ in that part of
the first term containing $\delta^2(k_1-k_3)$. This determines the relative
magnitude of the first two terms in $K^{(2)}_{2,2}$. We can conclude,
therefore, that the $O(g^2)$ Lipatov kernel is uniquely determined by
combining a knowledge of the reggeon diagrams that produce it with the
requirements of gauge invariance. We now go on to the $O(g^4)$ kernels,
which are uniquely determined in the same manner.

We must specifically consider the complete set of three loop diagrams. All
diagrams that contain three two reggeon states have already been included in
the sum of Fig.~7. That is, they are generated by two iterations of the
$O(g^2)$ kernel. The remaining diagrams are shown in Fig.~8. In these diagrams
all reggeon propagators corresponding to three and four reggeon states
are cancelled by nonsense zeroes and after construction by the multi-loop
procedure of \cite{arw3} each diagram reduces to the form (5) with a kernel
corresponding to one of the transverse momentum diagrams shown in Fig.~9.
The full kernel generated is therefore a symmetrised sum over terms with
such transverse momentum dependence. If we require both the Ward identity
constraint as $k_i \to 0, i=1,.,4$ and infra-red finiteness then the new
$O(g^4)$ part of the 2-2 kernel is uniquely determined to be
$$
\eqalign{g^{-4}& K^{(4)}_{2,2}(k_1,k_2,k_3,k_4)~=~
\sum_{\scriptscriptstyle 1<->2}
\Biggl(~{2 \over 3}(2\pi)^3 k_1^2J_2(k_1^2)k_2^2\Bigl(k_3^2\delta^2(k_2-k_4)
+~k_4^2\delta^2(k_2-k_3)\Bigr)~\cr
&-~\Biggl({k_1^2J_1(k_1^2)k_2^2k_3^2~+~k_1^2J_1(k_1^2)k_2^2k_4^2~+~
k_1^2k_3^2J_1(k_3^2)k_4^2~+~k_1^2k_3^2k_4^2J_1(k_4^2) \over
(k_1-k_3)^2} \Biggr)\cr
&+~J_1((k_1-k_3)^2)\Bigl(k_2^2k_3^2+k_1^2k_4^2\Bigr)
{}~+~k_1^2k_2^2k_3^2k_4^2~I(k_1,k_2,k_3,k_4)~\Biggr) }\auto\label{k422}
$$
where we have introduced
$$
\eqalign{ I(k_1,k_2,k_3,k_4)~=~{1 \over (2\pi)^3}\int d^2q {1 \over
q^2(q+k_1)^2(q+k_3)^2(q+k_1-k_4)^2}}\auto
$$
and
$$
\eqalign{ J_2(k^2)~~=~~{1 \over (2\pi)^3}\int d^2q {1 \over (k-q)^2}J_1(q^2)}
\auto
$$
The term involving $J_2$ in (\ref{k422}) is a three-particle contribution
to the reggeization of the gluon and has previously been found by
Bartels\cite{bar}. The term involving $I$ may also have been found in a
very recent paper by Bartels and Wusthoff\cite{bw}.

We now move on to the vertex which couples the color zero two-reggeon
state to the color zero four-reggeon state. We shall require that all
two-reggeon states within the four reggeon state also carry color zero. This is
a necessary requirement if we want to be sure we can write down a kernel
which is completely infra-red finite. There will also be contributions in
this kernel which contain a one-three reggeon vertex together with a
$\delta$-function in the remaining transverse momentum (analagous to the
first terms in both $K^{(2)}_{2,2}$ and $K^{(4)}_{2,2}$).

At $O(g^2)$ there is a one-three reggeon vertex\cite{arw3}, but it is a sum
of terms which necessarily carry color in a two-reggeon sub-channel and
contain either an overall or a sub-channel nonsense zero. These properties
allow this vertex to both produce the $O(g^4)$ three-particle state in the
trajectory function as we have discussed, and also to appear in the kernels
$K^{(4)}_{2,3}$ and $K^{(4)}_{2,4}$ that we introduce below. However, these
same properties prevent this vertex from directly coupling a three reggeon
cut to the one reggeon state. Such a vertex must carry even signature in
each two-reggeon sub-channel - for our purposes we can identify this
requirement with carrying zero-color in each sub-channel - and not have a
nonsense zero.

At $O(g^4)$ there is a one-three reggeon vertex generated from one-loop
diagrams which has the requisite properties. The reggeon diagrams involved
are shown in Fig.~10. The vertex can be expressed in the form
$$
\eqalign{ K^{(4)}_{1,3}(k,k_1,k_2,k_3)~=~{1 \over (2\pi)^3}\int
{d^2k_1 \over k_1^2}{d^2k_2 \over k_2^2}~k^2\delta^2(k-k_1-k_2)~
K^{(4)}_{2,3}(k_1,k_2,k_3,k_4,k_5)}\auto
$$
where $K^{(4)}_{2,3}(k_1,..,k_5)$ is a symmetrized two-three kernel
given by the sum of transverse momentum diagrams shown in Fig.~11. Since
this kernel can also be defined as a two-three reggeon amplitude it should
satisfy the Ward identity constraint. This uniquely determines the
coefficients of the various transverse momentum diagrams and gives
$$
\eqalign{ g^{-4}K^{(4)}_{2,3}&(k_1,k_2,k_3,k_4,k_5)~=~\sum_{\scriptscriptstyle
1<->2}~~\Biggl(~(k_1+k_2)^2 \cr
&-~\Biggl( {k_1^2(k_4+k_5)^2 \over (k_1-k_3)^2}
{}~+~{k_1^2(k_3+k_5)^2 \over (k_1-k_4)^2}
{}~+~{k_1^2(k_3+k_4)^2 \over (k_1-k_5)^2} \Biggr)\cr
&~+~{1 \over 3}\Biggl( {k_1^2k_5^2 \over (k_2-k_5)^2}
{}~+~{k_1^2k_4^2 \over (k_2-k_4)^2}
{}~+~{k_1^2k_3^2 \over (k_2-k_3)^2} \Biggr)\cr
&~+~{2 \over 3}\Biggl( {k_1^2k_2^2k_4^2 \over (k_1-k_3)^2(k_2-k_5)^2}
{}~+~{k_1^2k_2^2k_5^2 \over (k_1-k_4)^2(k_2-k_3)^2}
{}~+~{k_1^2k_2^2k_3^2 \over (k_1-k_5)^2(k_2-k_4)^2} \Biggr)\Biggr)}
\auto\label{k423}
$$
It is straightforward, if a little laborious, to check that
$$
\eqalign { K^{(4)}_{2,3}(k_1,k_2,k_3,k_4,k_5)
\centerunder{$\longrightarrow$} {\raisebox{-5mm} {$k_i \to 0$}}
{}~~0~~~i=1,.,5} \auto
$$
and so the gauge invariance constraint is indeed satisfied.

The complete two-four kernel is given by
$$
\eqalign{& K^{(4)}_{2,4}(k_1,k_2,k_3,k_4,k_5,k_6)~=~
{}~\sum_{\scriptscriptstyle 1<->2}
2\pi^3k_2^2\Biggl(\delta^2(k_2-k_6)K^{(4)}_{1,3}(k_1,k_3,k_4,k_5)\cr
&~+~\delta^2(k_2-k_5)K^{(4)}_{1,3}(k_1,k_3,k_4,k_6)
{}~+~\delta^2(k_2-k_4)K^{(4)}_{1,3}(k_1,k_3,k_5,k_6)\cr
&~+~\delta^2(k_2-k_3)K^{(4)}_{1,3}(k_1,k_4,k_5,k_6)\Biggr)
-~K^{(4)}_{2,4}(k_1,k_2,k_3,k_4,k_5,k_6)_c}
\auto\label{k424}
$$
The connected part of this kernel,
$K^{(4)}_{2,4}(k_1,..,k_6)_c$,
is given by the reggeon diagrams shown in Fig.~12 and the transverse
momentum diagrams to which they reduce are shown in Fig.~13. The symmetrised
sum over the distinct forms of transverse momentum dependence these diagrams
describe then gives
$$
\eqalign {& g^{-4}K^{(4)}_{2,4}(k_1,k_2,k_3,k_4,k_5,k_6)_c~=~
\sum_{\scriptscriptstyle 1<->2}~\Biggl(~(k_1 + k_2)^2
- \Biggl( {k_1^2(k_4+k_5+k_6)^2 \over (k_1-k_3)^2}\cr
&~+~{k_1^2(k_3+k_5+k_6)^2 \over (k_1-k_4)^2}
{}~+~{k_1^2(k_3+k_4+k_6)^2 \over (k_1-k_5)^2}
{}~+~{k_1^2(k_3+k_4+k_5)^2 \over (k_1-k_6)^2}\Biggr)
{}~-~{1 \over 4}\Biggl( {k_1^2k_3^2 \over (k_2-k_3)^2}\cr
&~+~{k_1^2k_4^2 \over (k_2-k_4)^2}
{}~+~{k_1^2k_5^2 \over (k_2-k_5)^2}
{}~+~{k_1^2k_6^2 \over (k_2-k_6)^2}\Biggr)
{}~+~{1 \over 2}\Biggl( {k_1^2(k_5+k_6)^2 \over (k_2-k_5-k_6)^2}
{}~+~{k_1^2(k_5+k_4)^2 \over (k_2-k_5-k_4)^2}\cr
&~+~{k_1^2(k_4+k_6)^2 \over (k_2-k_4-k_6)^2}
{}~+~{k_1^2(k_3+k_6)^2 \over (k_2-k_3-k_6)^2}
{}~+~{k_1^2(k_5+k_3)^2 \over (k_2-k_5-k_3)^2}
{}~+~{k_1^2(k_3+k_4)^2 \over (k_2-k_3-k_4)^2}\Biggr)\cr
&~+~{1 \over 2}\Biggl( {k_1^2k_2^2(k_4+k_5)^2 \over (k_1-k_3)^2(k_2-k_6)^2}
{}~+~{k_1^2k_2^2(k_3+k_5)^2 \over (k_1-k_4)^2(k_2-k_6)^2}
{}~+~{k_1^2k_2^2(k_3+k_4)^2 \over (k_1-k_5)^2(k_2-k_6)^2}\cr
&~+~{k_1^2k_2^2(k_3+k_6)^2 \over (k_1-k_4)^2(k_2-k_5)^2}
{}~+~{k_1^2k_2^2(k_4+k_6)^2 \over (k_1-k_3)^2(k_2-k_5)^2}
{}~+~{k_1^2k_2^2(k_5+k_6)^2 \over (k_1-k_3)^2(k_2-k_4)^2}\Biggr)\cr
&~-~{1 \over 4}\Biggl( {k_1^2k_2^2k_4^2 \over (k_1-k_3)^2(k_2-k_5-k_6)^2}
{}~+~{k_1^2k_2^2k_5^2 \over (k_1-k_3)^2(k_2-k_4-k_6)^2}
{}~+~{k_1^2k_2^2k_6^2 \over (k_1-k_3)^2(k_2-k_4-k_5)^2}\cr
&~+~{k_1^2k_2^2k_3^2 \over (k_1-k_4)^2(k_2-k_5-k_6)^2}
{}~+~{k_1^2k_2^2k_5^2 \over (k_1-k_4)^2(k_2-k_3-k_6)^2}
{}~+~{k_1^2k_2^2k_6^2 \over (k_1-k_4)^2(k_2-k_3-k_5)^2}\cr
&~+~{k_1^2k_2^2k_4^2 \over (k_1-k_5)^2(k_2-k_3-k_6)^2}
{}~+~{k_1^2k_2^2k_3^2 \over (k_1-k_5)^2(k_2-k_4-k_6)^2}
{}~+~{k_1^2k_2^2k_6^2 \over (k_1-k_5)^2(k_2-k_4-k_3)^2}\cr
&~+~{k_1^2k_2^2k_4^2 \over (k_1-k_6)^2(k_2-k_5-k_3)^2}
{}~+~{k_1^2k_2^2k_5^2 \over (k_1-k_6)^2(k_2-k_4-k_3)^2}
{}~+~{k_1^2k_2^2k_3^2 \over (k_1-k_6)^2(k_2-k_4-k_5)^2} \Biggr)\Biggr)}
\auto\label{k424c}
$$
The coefficients in (\ref{k424c}) are uniquely determined by requiring
first the vanishing when $k_i \to 0,~ i=1,.,6$ and in addition that
poles in the fourth and sixth terms cancel. The residues of these poles
involve a product of one-three vertices which can not be canceled (after
integration) by the $\delta$-function terms in (\ref{k424}) and so must
cancel within $K^{(4)}_{2,4}(k_1,..,k_6)_c$. Finally the
relative magnitude of the terms in (\ref{k424}) is determined by the
requirement that the remaining poles in
$K^{(4)}_{2,4}(k_1,..,k_6)_c$ do cancel with the
$\delta$-function terms to give an infra-red finite (integral) kernel.

All but the third term of $K^{(4)}_{2,3}(k_1,..,k_5)$ and the first, second
and fifth terms of $K^{(4)}_{2,4}(k_1,..,k_6)_c$ are already present in the
elementary kernels utilised by Bartels\cite{bar}. Also, in the calculation of
deep-inelastic high-mass diffraction by Bartels and Wusthoff\cite{bw} all
relevant additional terms are generated by application of the integral
equations of \cite{bar}. In fact, it is not straightforward to determine how
our kernel should contribute in this process since only some of the terms
clearly have the triple discontinuity that gives the leading-logs\cite{bw}.
(This issue needs more study before a direct comparison with the results of
\cite{bw} can be made). For example, the third term in
$K^{(4)}_{2,4}(k_1,..,k_6)_c$ arises from the third up to eighth reggeon
diagrams in Fig.~12, none of which has the relevant triple discontinuity.
These diagrams involve intermediate state reggeons for which neither the
reggeon propagator nor the signature factor pole survive the construction
procedure and also higher-order one-three and one-four reggeon couplings. In
Bartels' iteration of unitarity,these diagrams are generated by production
amplitude discontinuities which emerge after the (leading-log)
identification of the kernels. This is, of course, closely related to the
fact that they do not contribute, at leading log, in deep-inelastic
high-mass diffraction.

We will comment further on the utilisation of the kernels in \cite{arw3}.
 $K^{(4)}_{2,2}$ can clearly be added to $K^{(2)}_{2,2}$, the familiar
leading-order kernel, and the combination substituted in the Lipatov
equation. This would sum contributions arbitrarily far down in non-leading
logs and could give interesting indications of the significance of
higher-order corrections. However, since $K^{(4)}_{2,4}$ is also $O(g^4)$ it
can not be ignored in such an approximation. We could treat the combination
of $K^{(4)}_{2,2}$ and $K^{(4)}_{2,4}$ as a matrix operator between two and
four gluon states and write a closed set of coupled integral equations
involving only these states. Again since $K^{(4)}_{2,4}$ generates a
branching process which eventually couples to all even number gluon states
this would also be inconsistent. Indeed once $K^{(4)}_{2,4}$ is generated it
is hard to ignore the reality that a complete {\it Reggeon Field Theory} of
reggeon diagrams is being generated and that a solution in terms of a
physical Pomeron should perhaps be sought\cite{arw2}.

We conclude by expressing the hope that our formalism will ultimately
provide a direct construction procedure for writing down ``Yang-Mills reggeon
theories''. Since the defining Ward Identities of the theory are directly
imposed at the reggeon diagram level it seems possible that the underlying
Feynman diagram perturbation expansion can be by-passed completely in the
construction.

\newpage

\noindent{\bf Figure Captions}

\begin{itemize}

\item[{Fig.~1}] a) The identity between reggeon and zero momentum gluon
amplitudes. b) The relevant Ward identity.

\item[{Fig.~2}] The cut self-energy diagrams

\item[{Fig.~3}] Two loop reggeon diagrams.

\item[{Fig.~4}] Triple discontinuities and the corresponding transverse
momentum diagrams.

\item[{Fig.~5}] Two loop reggeon diagrams for the even signature amplitude.

\item[{Fig.~6}] Transverse momentum diagrams corresponding to the Lipatov
kernel.

\item[{Fig.~7}] The iteration of reggeon diagrams giving the Lipatov
equation.

\item[{Fig.~8}] Even signature three loop reggeon diagrams not included in
the iteration of the Lipatov kernel.

\item[{Fig.~9}] Transverse momentum diagrams describing the $O(g^4)$ kernel
generated by the reggeon diagrams of Fig.~8.

\item[{Fig.~10}] Reggeon diagrams contributing to the one-three reggeon
vertex.

\item[{Fig.~11}] Transverse momentum diagrams giving the $K^{(4)}_{2,3}$
kernel.

\item[{Fig.~12}] Reggeon diagrams contributing to the connected part of the
$K^{(4)}_{2,4}$ kernel.

\item[{Fig.~13}] Transverse momentum diagrams generated by the reggeon
diagrams of Fig.~12.

\end{itemize}

\end{document}